\documentstyle[psfig]{mn}
\def\degr{\hbox{$^\circ$}}
\def\>{$>$}
\def\<{$<$}
\def\simlt{\lower.5ex\hbox{$\; \buildrel < \over \sim \;$}}
\def\simgt{\lower.5ex\hbox{$\; \buildrel > \over \sim \;$}}

\newif\ifAMStwofonts

\ifoldfss
  \ifCUPmtlplainloaded \else
    \NewTextAlphabet{textbfit} {cmbxti10} {}
    \NewTextAlphabet{textbfss} {cmssbx10} {}
    \NewMathAlphabet{mathbfit} {cmbxti10} {} 
    \NewMathAlphabet{mathbfss} {cmssbx10} {} 
  \fi
  \ifAMStwofonts
    \ifCUPmtlplainloaded \else
      \NewSymbolFont{upmath} {eurm10}
      \NewSymbolFont{AMSa} {msam10}
      \NewMathSymbol{\upi}     {0}{upmath}{19}
      \NewMathSymbol{\umu}     {0}{upmath}{16}
      \NewMathSymbol{\upartial}{0}{upmath}{40}
      \NewMathSymbol{\leqslant}{3}{AMSa}{36}
      \NewMathSymbol{\geqslant}{3}{AMSa}{3E}

    \fi
  \fi
\fi 

\ifnfssone
  \newmathalphabet{\mathit}
  \addtoversion{normal}{\mathit}{cmr}{m}{it}
  \addtoversion{bold}{\mathit}{cmr}{bx}{it}
  \newmathalphabet{\mathbfit} 
  \addtoversion{normal}{\mathbfit}{cmr}{bx}{it}
  \addtoversion{bold}{\mathbfit}{cmr}{bx}{it}
  \newmathalphabet{\mathbfss} 
  \addtoversion{normal}{\mathbfss}{cmss}{bx}{n}
  \addtoversion{bold}{\mathbfss}{cmss}{bx}{n}
  \ifAMStwofonts
    \ifCUPmtlplainloaded \else
      \UseAMStwoboldmath
      \makeatletter
      \new@mathgroup\upmath@group
      \define@mathgroup\mv@normal\upmath@group{eur}{m}{n}
      \define@mathgroup\mv@bold\upmath@group{eur}{b}{n}
      \edef\UPM{\hexnumber\upmath@group}
      \new@mathgroup\amsa@group
      \define@mathgroup\mv@normal\amsa@group{msa}{m}{n}
      \define@mathgroup\mv@bold\amsa@group{msa}{m}{n}
      \edef\AMSa{\hexnumber\amsa@group}
      \makeatother
      \mathchardef\upi="0\UPM19
      \mathchardef\umu="0\UPM16
      \mathchardef\upartial="0\UPM40
      \mathchardef\leqslant="3\AMSa36
      \mathchardef\geqslant="3\AMSa3E
    \fi
  \fi
\fi 

\ifnfsstwo
  \DeclareMathAlphabet{\mathbfit}{OT1}{cmr}{bx}{it}
  \SetMathAlphabet\mathbfit{bold}{OT1}{cmr}{bx}{it}
  \DeclareMathAlphabet{\mathbfss}{OT1}{cmss}{bx}{n}
  \SetMathAlphabet\mathbfss{bold}{OT1}{cmss}{bx}{n}
  \ifAMStwofonts
    \ifCUPmtlplainloaded \else
      \DeclareSymbolFont{UPM}{U}{eur}{m}{n}
      \SetSymbolFont{UPM}{bold}{U}{eur}{b}{n}
      \DeclareSymbolFont{AMSa}{U}{msa}{m}{n}
      \DeclareMathSymbol{\upi}{0}{UPM}{"19}
      \DeclareMathSymbol{\umu}{0}{UPM}{"16}
      \DeclareMathSymbol{\upartial}{0}{UPM}{"40}
      \DeclareMathSymbol{\leqslant}{3}{AMSa}{"36}
      \DeclareMathSymbol{\geqslant}{3}{AMSa}{"3E}
    \fi
  \fi
\fi 

\ifCUPmtlplainloaded \else
  \ifAMStwofonts \else 
    \def\upi{\pi}
    \def\umu{\mu}
    \def\upartial{\partial}
  \fi
\fi

\pagerange {1--12}

\title{Eclipse studies of the dwarf-nova Ex Draconis}
\author[R. Baptista et~al.]
       {R. Baptista $^1$, M.\,S. Catal\'an $^2$ and L. Costa $^1$ \\
       $^1$ Departamento de F\'\i sica, Universidade Federal de Santa Catarina,
       Campus Trindade, 88040-900, Florian\'opolis - SC, Brazil, \\
       email: bap@fsc.ufsc.br \\
       $^2$ Department of Physics, Keele University, Keele, Staffordshire,
       ST5 5BG, UK, email: msc@astro.keele.ac.uk \\ }

\date{Submitted to MNRAS (1999 December 14)}

\pubyear{2000}

\begin{document}

\maketitle

\begin{abstract}

We report on $V$ and $R$ high speed photometry of the dwarf nova EX~Dra
in quiescence and in outburst. The analysis of the outburst lightcurves
indicates that the outbursts do not start in the outer disc regions.
The disc expands during the rise to maximum and shrinks during decline
and along the following quiescent period. The decrease in brightness at
the later stages of the outburst is due to the fading of the light from the
inner disc regions. At the end of two outbursts the system was seen to go
through a phase of lower brightness, characterized by an out-of-eclipse
level $\simeq 15$ per cent lower than the typical quiescent level and by
the fairly symmetric eclipse of a compact source at disc centre with
little evidence of a bright spot at disc rim.

New eclipse timings were measured from the lightcurves taken in quiescence
and a revised ephemeris was derived. The residuals with respect to the
linear ephemeris are well described by a sinusoid of amplitude 1.2 minutes
and period $\simeq 4$ years and are possibly related to a solar-like 
magnetic activity cycle in the secondary star. Eclipse phases of the
compact central source and of the bright spot were used to derive the
geometry of the binary. By constraining the gas stream trajectory to pass
through the observed position of the bright spot we find $q=0.72\pm 0.06$
and $i= 85^{+3}_{-2}$ degrees. The binary parameters were estimated by
combining the measured mass ratio with the assumption that the secondary
star obeys an empirical main sequence mass-radius relation. We find
$M_1 = 0.75\pm 0.15 \; M_\odot$ and $M_2 = 0.54\pm 0.10 \; M_\odot$.
The results indicate that the white dwarf at disc centre is surrounded
by an extended and variable atmosphere or boundary layer of at least 3
times its radius and a temperature of $T\simeq 28000 \;K$. The fluxes
at mid-eclipse yield an upper limit to the contribution of the secondary
star and lead to a lower limit photometric parallax distance of $D= 290
\pm 80\; pc$. The fluxes of the secondary star are well matched by those
of a M$0\pm2$ main sequence star.

\end{abstract}


\section{Introduction}

Dwarf novae are mass-exchanging binaries in which a late type star (the
secondary) overfills its Roche lobe and transfers matter to a companion
white dwarf (the primary) via an accretion disc. 
These systems show recurrent outbursts
of 2--5 magnitudes on timescales of a few weeks to months caused either
by an instability in the mass transfer from the secondary star or by a
thermal instability in the accretion disc which switches the disc from
a low to a high-viscosity regime (Warner 1995 and references therein).
During outburst most of the light arises from the bright, optically thick
accretion disc, while in quiescence the dominant sources of light are the
white dwarf and the bright spot formed by the impact of the infalling
gas stream with the edge of the disc.

Eclipsing dwarf novae are probably the best sites for the study of 
accretion physics as the occultation of the accretion disc and white
dwarf by the secondary can be used to constrain the geometry and 
parameters of the binary, and tomographic techniques such as eclipse
mapping (Horne 1985) and Doppler tomography (Marsh \& Horne 1988) can
be applied to probe the structure and dynamics of the accretion flow.

EX Draconis (= HS1804+67) was detected in the Hamburger Quasar Survey
(Bade et~al. 1989) and shown to be an eclipsing dwarf nova with an
orbital period of 5.04 hr by Barwig et~al. (1993). From spectroscopic
observations made in quiescence, Billington, Marsh \& Dhillon (1996)
found that the secondary star is of spectral type M1 to M2 and that
it contributes almost all of the light at mid-eclipse. Their analysis
showed that the inner face of the secondary is significantly irradiated
by the white dwarf. They found a rotational broadening of $v\sin i =
140\; km\,s^{-1}$ and a radial velocity semi-amplitude of $K_2 = 210\;
km\,s^{-1}$ for the secondary star which leads to a spectroscopic mass
ratio of $q=0.8$ when combined with the $K_1= 167\; km\,s^{-1}$ of Barwig
et~al. (1993). A relatively small value for the radius of the accretion
disc ($0.4\;R_{L1}$) is derived but no explanation is given of how this
estimate was made.

In a follow up study using spectroscopy and photometry of EX~Dra in
quiescence and in outburst, Fiedler, Barwig \& Mantel (1997) measured
radial velocity semi-amplitudes of $K_1= 167\; km\,s^{-1}$ and $K_2=
223\; km\,s^{-1}$ and derived a spectroscopic model for the binary with
$q=0.75$, $i= 84.2\degr$, $M_1= 0.75\;M_\odot$ and $M_2= 0.56\;M_\odot$.
However, the radial velocity curve of the H$\alpha$ line shows a large
phase shift ($\simeq 0.2$ cycle) with respect to photometric conjunction
which casts doubt on the derived value of $K_1$. They use the eclipse 
phases of the bright spot and white dwarf to derive a photometric mass
ratio between 0.7 and 0.8, supporting the spectroscopic model. From the
ratios of Ca\,I and TiO absorption features they infer a spectral type
of M0 for the secondary star. Smith \& Dhillon (1998) use the values of
$v\sin i$ and $K_2$ of Billington et~al. (1996) and the eclipse phase
width $\Delta\phi$ of Fiedler et~al. (1997) to infer a $K_1= 176 
\;km\,s^{-1}$.

In this paper we present and discuss high-speed photometry of EX~Dra in
quiescence and in outburst. Section~\ref{obs} describes the observations.
In section~\ref{results} we present and discuss the eclipse lightcurves,
provide an updated ephemeris, derive the binary parameters from the
eclipse phases of the white dwarf and bright spot, and obtain estimates
of the distance to the binary. The results are summarized in
section~\ref{final}.

\section{Observations and data reduction} \label{obs}

Time-series of differential photometry of EX~Dra in the $V$ and $R$ bands
was obtained with a Wright Instruments CCD camera (1.76 arcsec/pixel, 
$385 \times 578$ pixels) attached to the 0.9-m James Gregory Telescope
of the University Observatory, St.\,Andrews, in 1995 and 1996. This pixel
size is matched to the seeing at this sea-level site, where typical stellar
images have FWHM values of 3.0 pixels, and are therefore well sampled.
Exposure times ranged from 15 to 40\,s with a dead-time between exposures
of about 5\,s to read the CCD chip. Details of the observations are listed
in Table~\ref{tab1}. The observations include five outbursts of EX~Dra 
and sample various phases along the outburst cycle.
%
\begin{table*}
 \centering
 \begin{minipage}{140mm}
  \caption{Journal of the observations.} \label{tab1}
  \begin{tabular}{@{}lrcrccccl@{}}
~~~ Date & Start & $\Delta t$ & No. of ~~ & Band & Cycle & Phase range &
Quality \footnote{ A= photometric (main comparison stable), B= good (some
sky variations), C= poor (large variations and/or clouds).} & State \\ [-0.5ex]
& (UT) & (s) & exposures & & & (cycles) \\ [1ex]
1995 Sep 20 & 21:23 & 30 &  96~~~ & R & 7540 & $+0.06,+0.32$ & C & maximum \\
1995 Sep 24 & 20:36 & 30 & 113~~~ & R & 7559 & $-0.04,+0.19$ & A & decline \\
1995 Sep 25 & 21:28 & 30 & 103~~~ & R & 7564 & $-0.10,+0.18$ & A & decline \\
1995 Sep 26 & 22:45 & 30 &  97~~~ & R & 7569 & $-0.08,+0.12$ & B & quiescence\\
[0.5ex]
1995 Oct 15 & 20:32 & 30 & 101~~~ & R & 7659 & $-0.02,+0.19$ & B & maximum \\
1995 Oct 16 &  0:48 & 30 &  54~~~ & R & 7660 & $-0.17,-0.01$ & C & maximum \\
1995 Oct 17 & 22:07 & 30 & 178~~~ & R & 7669 & $-0.18,+0.19$ & B & decline \\
1995 Oct 18 &  3:34 & 30 &  71~~~ & R & 7670 & $-0.10,+0.09$ & C & decline \\
1995 Oct 18 & 18:41 & 30 &  74~~~ & R & 7673 & $-0.10,+0.04$ & B & decline \\
			&  0:17 & 30 &  78~~~ & R & 7674 & $+0.01,+0.16$ & C & decline \\
[0.5ex]
1995 Nov 14 &  1:20 & 30 & 156~~~ & R & 7798 & $+0.07,+0.52$ & B & quiescence\\
1995 Nov 16 & 22:24 & 30 & 149~~~ & R & 7812 & $-0.22,+0.19$ & C & rise \\
1995 Nov 17 &  3:39 & 30 & 113~~~ & R & 7813 & $-0.18,+0.12$ & B & rise \\
1995 Nov 18 &  0:21 & 30 &  82~~~ & R & 7817 & $-0.07,+0.09$ & B & rise \\
1995 Nov 19 & 21:37 & 30 &  53~~~ & R & 7826 & $-0.09,+0.12$ & C & maximum \\
1995 Nov 20 &  2:34 & 30 &  88~~~ & R & 7827 & $-0.11,+0.09$ & B & maximum \\
1995 Nov 24 & 22:59 & 30 &  49~~~ & R & 7850 & $-0.00,+0.10$ & C & decline \\
[0.5ex]
1995 Dec 11 &  2:04 & 15/18 & 252~~~ & R & 7927 & $-0.18,+0.13$ & A & rise \\
1995 Dec 20 &  3:01 & 15 & 220~~~ & R & 7970 & $-0.12,+0.13$ & A & decline \\
1995 Dec 20	& 22:59 & 30 & 160~~~ & R & 7974 & $-0.16,+0.18$ & A & quiescence\\ 
1995 Dec 26 & 20:14 & 30 & 172~~~ & R & 8002 & $-0.12,+0.23$ & A & low state \\
1995 Dec 27 &  1:12 & 30 & 182~~~ & R & 8003 & $-0.13,+0.24$ & A & low state \\
			& 6:19 & 30/40 & 91~~~ & V & 8004 & $-0.12,+0.07$ & A & low state\\
1995 Dec 27 & 21:31 & 30 & 148~~~ & R & 8007 & $-0.10,+0.19$ & A & quiescence\\
1995 Dec 28 &  2:10 & 30 & 188~~~ & R & 8008 & $-0.18,+0.21$ & A & quiescence\\
1995 Dec 28 & 17:46 & 30 & 177~~~ & R & 8011 & $-0.09,+0.26$ & A & quiescence\\
			& 22:25 & 30 & 188~~~ & V & 8012 & $-0.16,+0.25$ & A & quiescence\\
1995 Dec 29 &  3:20 & 30 & 202~~~ & R & 8013 & $-0.19,+0.22$ & A & quiescence\\
1995 Dec 29 & 18:13 & 30 & 214~~~ & R & 8016 & $-0.23,+0.20$ & B & quiescence\\
			& 23:43 & 30/40 & 132~~~ & R & 8017 & $-0.14,+0.14$ & C &
quiescence \\ [0.5ex]
1996 Jan 10 & 21:40 & 15-40 & 224~~~ & R & 8074 & $-0.39,+0.28$ & B &
quiescence \\ [0.5ex]
1996 Nov 22	& 18:00 & 30 & 152~~~ & R & 9583 & $-0.14,+0.18$ & B & decline \\
[-3ex]
\end{tabular}
\end{minipage}
\end{table*}

The data was reduced using standard {\sc IRAF}\footnote{ 
IRAF is distributed by National Optical Astronomy Observatories,
which is operated by the Association of Universities for Research in
Astronomy, Inc., under contract with the National Science Foundation.}
procedures and included bias and flat-field corrections and cosmic rays
removal. Photometry was obtained with the automated aperture photometry
routine JGTPHOT (Bell, Hilditch \& Edwin 1993). Fluxes were extracted for
the variable and for five selected comparison stars in the field. The
relative brightness of the comparison stars in all data sets is constant
to better than 0.01 mag. We adopted a mean comparison star magnitude for
each frame from the intensity-added values of these five stars. Time-series
were constructed by computing the magnitude difference between the variable
and the mean comparison star. From the dispersion in the magnitude 
difference of the comparison stars with similar brightness we estimate
an uncertainty in the photometry of EX~Dra of 0.025 mag in quiescence 
and better than 0.01~mag in outburst.

Observations of spectrophotometric standard stars of Massey et~al. (1998)
and Massey \& Gronwall (1990) were used to calibrate the photometry in
the EX~Dra field. These observations demonstrate that the transformation
coefficients from the natural to the standard $VR$ system (Bessell 1983)
are unity to a precision of one per cent. Hence differential instrumental
magnitudes obtained from individual frames are differential $V$ and $R$
magnitudes. We found the mean of the combined $V$ and $R$ magnitudes of
the five comparison stars to be $V= 13.21 \pm 0.09$ mag and $R= 12.84
\pm 0.05$ mag. We used the relations $V= 16.40 - 2.5\; \log_{10}f_\nu
[{\rm mJy}]$ and $R= 16.22 - 2.5\; \log_{10}f_\nu [{\rm mJy}]$ (Lamla 1982)
to transform the calibrated magnitudes to absolute flux units.

\section{Results} \label{results}

\subsection {Eclipse lightcurves} \label{cluz}

We adopted the following convention regarding the phases: conjunction
occurs at phase zero, the phases are negative before conjunction and
positive afterwards. The lightcurves were phased according to the
ephemeris of eq.\,\ref{efem} (section~\ref{rev_efem}).

\begin{figure*}
\centerline{\psfig{figure=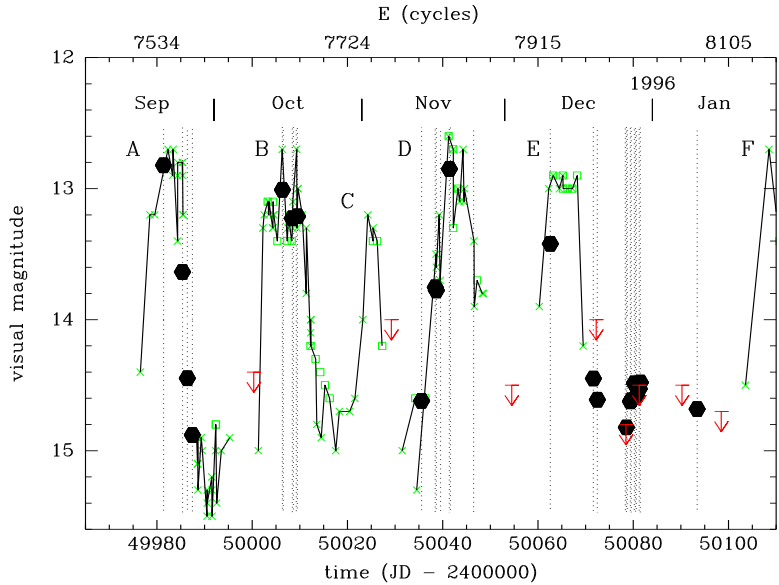,angle=-90,width=12cm,rheight=8cm}}
 \caption{ Visual lightcurve of EX~Dra during the period September 1995
 to January 1996, constructed from observations made by the AAVSO (crosses)
 and VSNET (open squares). Arrows indicate upper limits on the visual
 magnitude. Vertical dotted lines mark the epochs of our observations.
 $R$-band out-of-eclipse magnitudes from our dataset are shown as filled
 circles for illustration purposes.} 
\label{fig1}
\end{figure*}

Figure~\ref{fig1} shows the visual lightcurve of EX~Dra for the period
September 1995 to January 1996 from AAVSO and VSNET observations.
%
Vertical dotted lines mark the epochs of our observations while filled
circles show the corresponding $R$-band out-of-eclipse magnitudes.
There were six recorded outbursts during this period (labeled from
A to F in Fig.\,\ref{fig1}), with typical amplitudes of $\simeq 2.0$ mag,
duration of $\simeq 10$ days, and average time between outbursts of
$20\pm 3$ days. Outburst C was shorter ($\simeq 5$ days) and weaker
($\Delta m\simeq 1.5$ mag) than the others and, unfortunately, was not
covered by our observations. The visual magnitude is typically $m_v
\simeq 12.7$ at maximum and $m_v\simeq 15$ in quiescence. At the end
of outbursts A and E the star went through a faint state (hereafter 
named low state) before recovering its usual quiescent level.

Our dataset frames eclipse lightcurves during most relevant phases
through the outburst cycle: early rise to maximum (cycles 7812, 7813,
7817), late rise to maximum (7927), maximum light (7540, 7659, 7660,
7826 and 7827), end of maximum (7669, 7670, 7673, 7674 and 9583),
early decline (7559, 7850), late decline (7564, 7970), low state
(8002 to 8004), and quiescence (7569, 7798, 8007 to 8017, and 8074).

\begin{figure}
\centerline{\psfig{figure=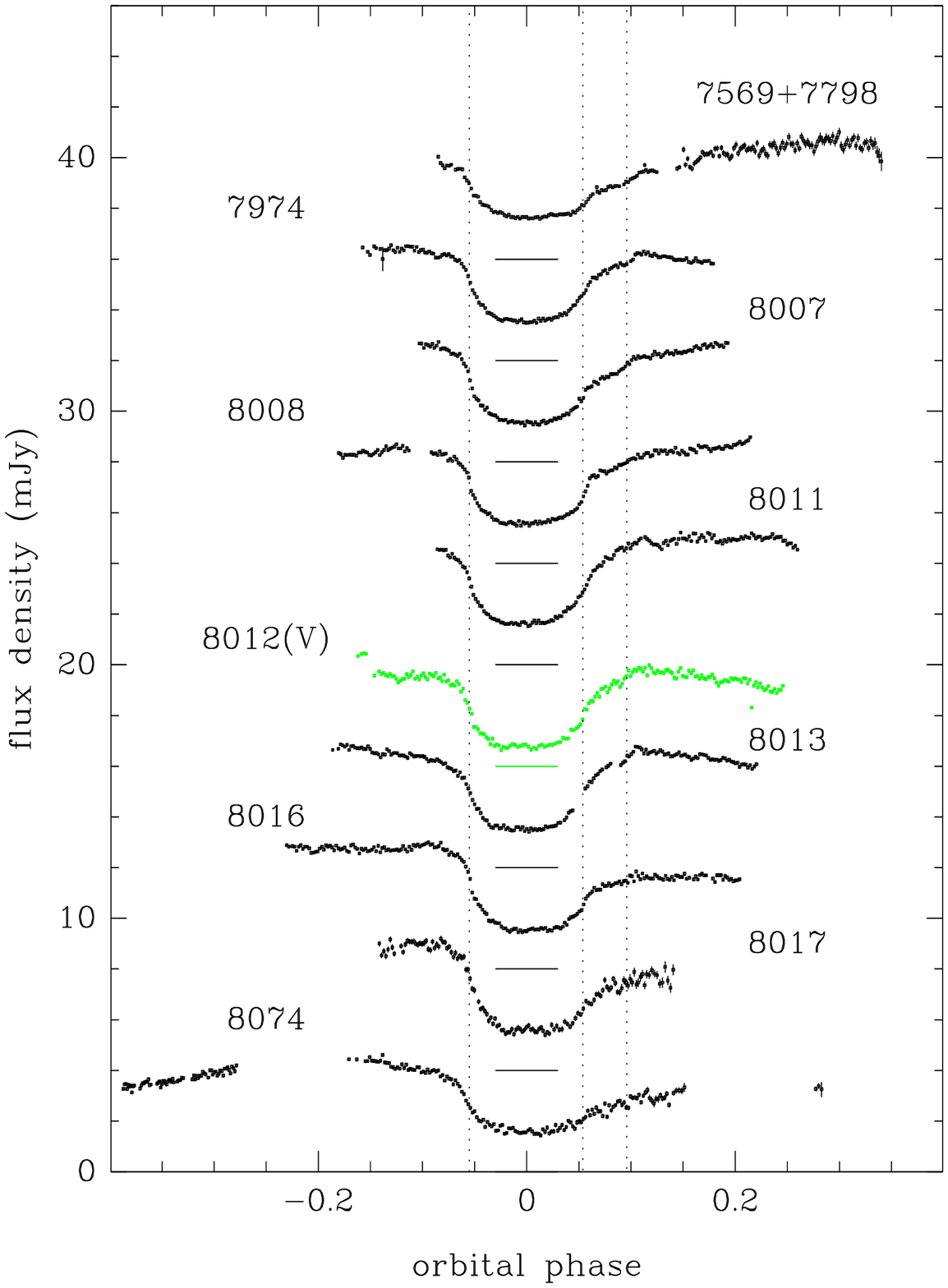,width=10.3cm,rheight=13cm}}
 \caption{ Lightcurves of EX~Dra in quiescence. The curves are 
  progressively displaced upwards by 4~mJy. Horizontal lines at 
  mid-eclipse show the true zero level in each case. Vertical dotted
  lines mark the phases of ingress/egress of the bright spot and the
  egress of the white dwarf as measured in section~\ref{param}. 
  Labels indicate the cycle number. } 
\label{fig2}
\end{figure}

Individual lightcurves of EX~Dra in quiescence are shown in 
Fig.~\ref{fig2}. 
%
Sharp changes in the slope reflect the occultation of
a compact source at disc centre and of the bright spot at disc rim. The
ingress of the central source and of the bright spot overlap in phase
to form a unique sharp break in the slope. The egress of the central
source is variable both in duration and in flux and sometimes is hardly
visible (e.g., cycle 8074). The eclipses have a flat-bottomed, `U'
shape indicating that the eclipse is close from being total. There is
no pronounced flickering (amplitude $\simlt 2.5$ per cent). In some of
the lightcurves the flux level is the same before and after eclipse
while others show a perceptible orbital hump prior to eclipse (usually
interpreted as being the result of anisotropic emission from the bright
spot) and a slow recovery from eclipse where the bright spot egress
is hardly discernible.

\begin{figure*}
\centerline{\psfig{figure=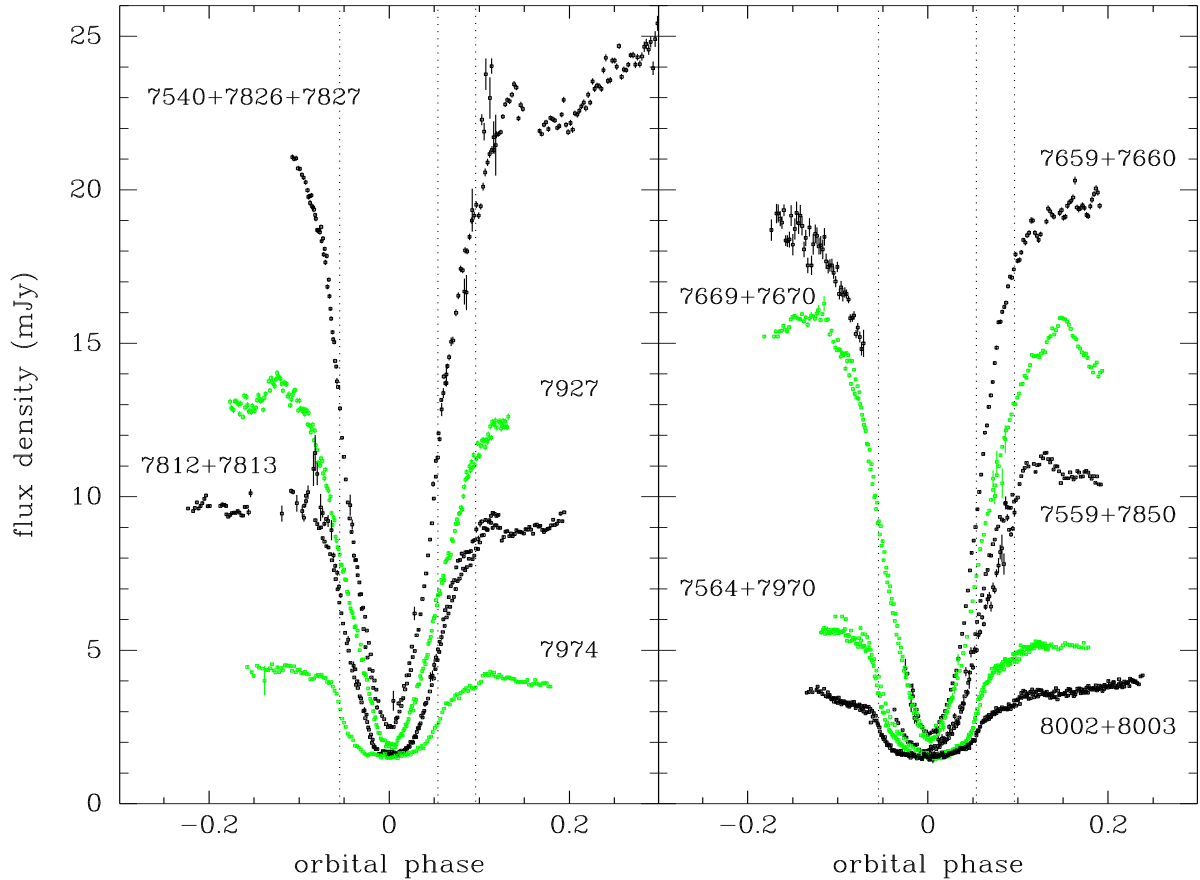,angle=-90,width=18.5cm,rheight=13cm}}
 \caption{ Lightcurves of EX~Dra through the outburst cycle. Vertical 
  dotted lines mark the phases of ingress/egress of the bright spot and
  the egress of the white dwarf as measured in section~\ref{param}. 
  Labels indicate the cycle number. The left panel shows lightcurves on
  the rise to maximum and at maximum while the right panel shows lightcurves
  during decline and in the low state. The quiescent lightcurve of cycle
  7974 is shown in the left panel for reference. } 
\label{fig3}
\end{figure*}

Figure~\ref{fig3} shows lightcurves of EX~Dra along the outburst cycle.
The left panel shows the behaviour during rise and maximum light while
the right panel shows the behaviour during decline and in the low state.
The lightcurves were grouped by outburst phase. Only a subset of the
outburst lightcurves are shown for clarity. The fact that, for a given
outburst stage, the lightcurves of different outbursts have similar 
eclipse shapes and out-of-eclipse levels gives confidence that the
observed sequence is representative of the general behaviour of EX~Dra
during outburst (at least for the epoch of our observations).
%

Lightcurve 7812+7813 frames the early rise to maximum. The eclipse shape
is asymmetric and mid-eclipse occurs after phase zero, indicating that
the receding side of the disc is brighter. The mid-eclipse level and the
total eclipse width are the same as in quiescence, showing that the
brightening does not start in the outer disc regions. This is in agreement
with the symmetric shape of the outbursts, with comparable rise and 
decline timescales (cf. Fig.\,\ref{fig1}), which is typical of inside-out
outbursts (e.g., Cannizzo, Wheeler \& Polidan 1986).

The eclipse profile changes during the rise to maximum, from the 
asymmetric `U' shape eclipse of a compact central source plus the bright
spot at disc rim to a more symmetric `V' shape indicating the partial
occultation of a bright extended disc. The total width of the eclipse
increases during the rise (from 0.196 cycle in quiescence to about 0.22
cycle at lightcurve 7927 and larger at maximum) indicating that the
disc radius also increases as the system approaches maximum light. A
precise measurement of the total width of the eclipse at maximum light
is precluded due to the limited phase coverage of the corresponding
lightcurve.

The disc shrinks during decline (as indicated by the eclipse egress
phase) until it reaches the quiescent radius close to the end of the
outburst (lightcurve 7564+7970). The low state is characterized by an
out-of-eclipse level $\simeq 15$ per cent lower than the typical
quiescent level and by a fairly symmetric eclipse shape, corresponding
to the eclipse of a compact source at the disc centre with little 
evidence of a bright spot at the disc rim. The mid-eclipse level of
lightcurves 7564+7970 and 8002+8003 is the same, showing that the
decrease in brightness at this stage is due to the fading of the light
from the inner disc regions.

Flickering is much more pronounced in outburst than in quiescence.
Flickers of an amplitude of $\simeq 10-15$ per cent can be seen in
many lightcurves in outburst.

None of our outburst lightcurves resembles the flat-bottomed outburst
lightcurve of Fiedler et~al. (1997, see their fig.~6). Their outburst
lightcurve is a factor of only $\simeq 2$ brighter than their typical
quiescent lightcurve indicating that it corresponds to outbursts of
lower amplitude than the ones sampled by our observations.

\subsection {Revised ephemeris} \label{rev_efem}

The ingress feature of the white dwarf and of the bright spot overlap
in quiescent lightcurves of EX~Dra (Fig.~\ref{fig2}). Since the bright
spot ingress depends on the variable disc radius, the mid-ingress time
is not a stable feature of the eclipse. We therefore adopted the same
procedure of Fiedler et~al. (1997) and used the mid-egress times of the
white dwarf plus the inferred duration of its eclipse (see 
section~\ref{param}) to obtain a revised ephemeris for the mid-eclipse
times.

Mid-egress times were measured by computing the time of maximum derivative
in a median-filtered version of the lightcurve (section~\ref{param}). 
The uncertainty in determining mid-egress times depends on the time
resolution and signal-to-noise of the lightcurve and is in the range
$(1-2) \times 10^{-4}$~d. The barycentric correction and the difference
between universal times (UT) and dynamical ephemeris time scales are 
smaller than the uncertainties in the measured timings and were neglected.
The new heliocentric (HJD) times of the egress of the white dwarf are
listed in Table~\ref{tab2} with corresponding cycle number and 
uncertainties (quoted in parenthesis).
%
\begin{table}
 \centering
 \begin{minipage}{50mm}
  \caption{New eclipse timings.} \label{tab2}
  \begin{tabular}{@{}ccl@{}}
cycle & white dwarf egress & (O$-$C) \footnote{Observed minus Calculated
times with respect to the linear part of the ephemeris of eq.~\ref{efem}.} 
\\ [-0.5ex]
 & HJD (2\,400\,000 +) & ~(cycle) \\ [1ex]
7569 & 49987.4778 (2) & $+0.0009$ \\
7974 & 50072.5023 (2) & $+0.0014$ \\
8002 & 50078.3799 (2) & $-0.0017$ \\
8003 & 50078.5900 (1) & $-0.0010$ \\
8004 & 50078.8000 (1) & $-0.0006$ \\
8007 & 50079.4296 (2) & $-0.0017$ \\
8008 & 50079.6404 (1) & $+0.0025$ \\
8011 & 50080.2700 (2) & $+0.0015$ \\
8012 & 50080.4793 (1) & $-0.0016$ \\
8013 & 50080.6893 (1) & $-0.0015$ \\
8016 & 50081.3191 (1) & $-0.0015$ \\
8017 & 50081.5290 (1) & $-0.0015$ \\
8074 & 50093.4955 (1) & $-0.0010$ \\ [-3ex]
\end{tabular}
\end{minipage}
\end{table}

\begin{figure}
\centerline{\psfig{figure=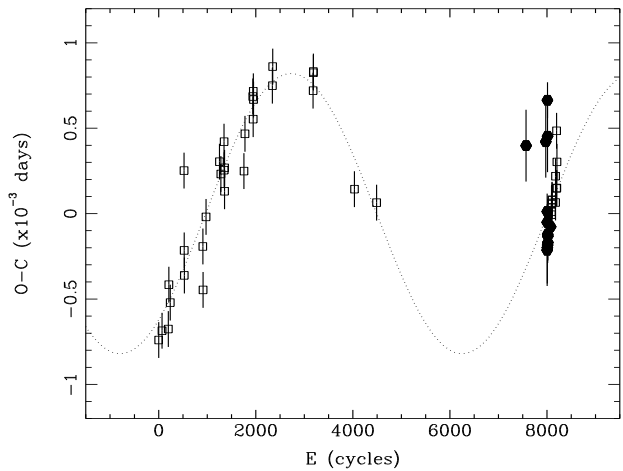,angle=-90,width=9.5cm,rheight=6.8cm}}
 \caption{ The $(O-C)$ diagram for the eclipse mid-egress times calculated
 from the linear part of eq.(\ref{efem}). The timings of Fiedler et~al.
 (1997) are plotted as open squares while those in Table~\ref{tab2} are
 shown as filled circles. } 
\label{fig4}
\end{figure}

We assumed equal errors of $10^{-4}$~d for the timings of Fiedler et~al.
(1997) and combined them with the timings of Table~\ref{tab2} to obtain
a least-squares linear fit with a reduced $\chi^2= 13.8$ for 54 d.o.f.
and a standard deviation of $\sigma= 0.0019$ cycle. The residuals with
respect to the linear ephemeris show a clear cyclical behaviour and can
be well described by a sinusoidal function. Assuming a duration of the
eclipse of the white dwarf of $\Delta t=0.0228$~d (section~\ref{param}),
the best-fit linear plus sinusoidal ephemeris for the mid-eclipse times
is,
\[
T_{mid} = {\rm HJD}\; 2\,448\,398.4530(\pm 1) + 0.209\,936\,98(\pm 4)\,E +
\]
\begin{equation}
+ (8.2 \pm 1.5) \times 10^{-4} \; \sin \left[ 2\pi \frac{(E-968)}{7045}
\right] \; d \;\;\; ,
\label{efem}
\end{equation}
with $\chi^2= 2.7$ for 51 d.o.f. and $\sigma= 0.0010$ cycle.
Residuals with respect to the linear part of eq.(\ref{efem}) are listed
in Table~\ref{tab2} and shown in Fig.~\ref{fig4}. The sinusoidal term
of eq.(\ref{efem}) is indicated in Fig.~\ref{fig4} as a dotted line.
%

The amplitude (1.18 minutes) and timescale ($\simeq 4$ years) of the
period variation are similar to the quasi-periodic orbital period changes
found in many other eclipsing CVs (Warner 1995 and references therein)
and are possibly related to a solar-like magnetic activity cycle in
the secondary star (Applegate 1992; Richman, Applegate \& Patterson 
1994). It may also be possible that the period changes are due to a
third body in the system, as suggested by Fiedler et~al. (1997), if the
variation proves to be strictly periodic. Regular observations of eclipse
timings during the next decade are required in order to check the 
stability of the period of the variation and test the above hypotheses.

\subsection {Binary parameters} \label{param}

\subsubsection {Measuring eclipse phases} \label{phases}

The ingress/egress phases of the occultation of the compact central
source (hereafter CS) and of the bright spot (BS) by the secondary star
provide information about the geometry of the binary system and the
relative sizes of these components (e.g., Wood et~al. 1986).

We used the lightcurves of the low state -- where the effect of the BS
is minimal on the eclipse shape -- to measure the ingress and egress
phases of the CS and to derive the width of its eclipse as well as the
duration of its ingress/egress feature. The contact phases can be 
identified as rapid changes in slope visible in the lightcurves of
the low state (Fig.~\ref{fig3}) and were measured with the aid of a
cursor on a graphic display of a median filtered version of the
lightcurve. We defined $\phi_{c1},\phi_{c2}$ as those phases during
which the CS disappears behind the secondary star and $\phi_{c3},
\phi_{c4}$ as the phases corresponding to its reappearance from eclipse.
The mid-ingress (egress) phases ($\phi_{ci},\phi_{ce}$) were computed
as the phases at which half of the central source light is eclipsed
and also as the phases of minimum (maximum) derivative in the lightcurve
(e.g., Wood, Irwin \& Pringle 1985).

\begin{figure}
\centerline{\psfig{figure=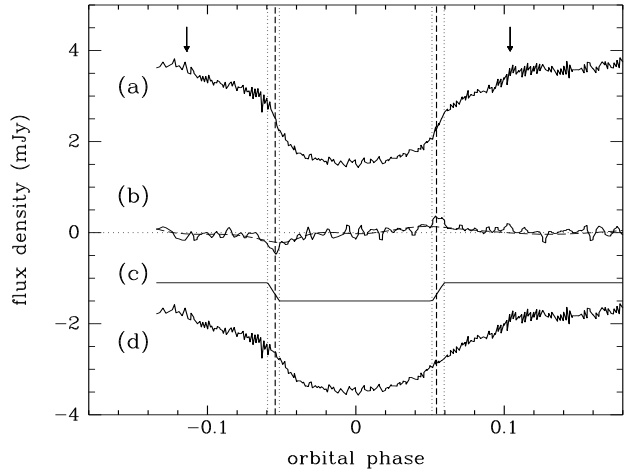,angle=-90,width=9.5cm,rheight=7cm}}
 \caption{ Measuring the eclipse phases of the compact central source.
 (a) Lightcurve 8002+8003. (b) The median-filtered derivative of (a), 
 multiplied by a factor 2. The dashed line is a spline fit to the
 extended and slowly varying eclipse of the disc. (c) The reconstructed
 central source lightcurve, shifted downwards by 1.5 mJy. (d) Lightcurve
 (a) after subtraction of the central source component, shifted 
 downwards by 5 mJy. Dashed lines mark mid-ingress and mid-egress
 phases of the central source and dotted lines mark its four contact
 phases. Arrows indicate the beginning and end of the eclipse of the
 disc. } 
\label{fig5}
\end{figure}

Figure~\ref{fig5} illustrates the measurement procedure with the 
derivative for the combined lightcurve 8002+8003. The ingress/egress
of the CS can be seen as those intervals for which the derivative curve
is significantly different from zero. 
%
The width at half-peak intensity of these features yields a preliminary
estimate of their duration. A spline function is fitted to the remaining
regions in the derivative curve to remove the contribution from the
extended and slowly varying eclipse of the disc. Estimates of the CS
flux at ingress (egress) are obtained by integrating the flux in the
spline-subtracted derivative curve between the first and second (third
and fourth) contact phases. The lightcurve of CS is then reconstructed
by assuming that the flux is zero between ingress and egress and that it
is constant outside eclipse. The reconstructed CS lightcurve can be seen
in Fig.~\ref{fig5}(c) and the lightcurve after removal of the CS component
is shown in Fig.~\ref{fig5}(d).

The measured contact phases, mid-ingress and mid-egress phases of the CS
from the lightcurves of the low state are collected in Table~\ref{tab3}.
The quoted mid-ingress/egress values are the average of both procedures
described above and have an estimated error of 0.0005 cycle.
%
\begin{table*}
 \centering
 \begin{minipage}{163mm}
  \caption{Eclipse parameters from lightcurves of the low state.} \label{tab3}
  \begin{tabular}{@{}lcccccccrcc@{}}
cycle & $\phi_{c1}$ & $\phi_{c2}$ & $\phi_{c3}$ & $\phi_{c4}$ &
$\phi_{ci}$ & $\phi_{ce}$ & $\Delta\phi$ & $\phi_0\;\;\;$ &
$\Delta_{ci}$ & $\Delta_{ce}$ \\ [1ex]
8002 		& $-0.0595$ & $-0.0515$ & +0.0515 & +0.0595 &
$-0.0545$ & +0.0545 & 0.1090 & 0.0000 & 0.0080 & 0.0080 \\
8003 		& $-0.0575$ & $-0.0490$ & +0.0515 & +0.0600 &
$-0.0540$ & +0.0540 & 0.1080 & 0.0000 & 0.0085 & 0.0085 \\
8004V 		& $-0.0580$ & $-0.0500$ & +0.0515 & +0.0600 &
$-0.0540$ & +0.0550 & 0.1090 & +0.0005 & 0.0080 & 0.0085 \\
8002+8003	& $-0.0595$ & $-0.0515$ & +0.0515 & +0.0595 &
$-0.0540$ & +0.0540 & 0.1080 & 0.0000 & 0.0080 & 0.0080 \\ [0.5ex]
mean 		& $-0.0586$ & $-0.0505$ & +0.0515 & +0.0598 &
$-0.0541$ & +0.0544 & 0.1085 & +0.0001 & 0.0081 & 0.0082 \\
error 		& $\pm 0.0010$ & $\pm 0.0010$ & $\pm 0.0005$ & $\pm 0.0003$ &
$\pm 0.0003$ & $\pm 0.0005$ & $\pm 0.0006$ & $\pm 0.0003$ & $\pm 0.0003$ &
$\pm 0.0003$ \\
\end{tabular}
\end{minipage}
\end{table*}

The duration of the CS eclipse (the eclipse of the disc centre) is 
defined as
\begin{equation}
\Delta\phi= \phi_{ce} - \phi_{ci} \; ,
\end{equation}
and the mid-eclipse phase (the inferior conjunction of the binary) 
is written as
\begin{equation}
\phi_0 = 1/2\:(\phi_{ce} + \phi_{ci}) \; .
\end{equation}
These quantities are collected in Table~\ref{tab3}.
The mean of the measurements from all lightcurves yields $\Delta\phi =
0.1085 \pm 0.0006$ cycle (=0.0228 d), where the quoted error is the
standard deviation of the mean. Similarly, we have $\phi_0 = +0.0001
\pm 0.0003$ cycle, which indicates that the centre of the CS eclipse
corresponds to phase zero.
The difference between the first and second (third and fourth) CS
contact phases yield the phase width of the CS ingress (egress),
$\Delta_{\rm ci}$ ($\Delta_{\rm ce}$). These quantities are also listed
in Table~\ref{tab3}.  A mean from all values of $\Delta_{\rm ci}$
and $\Delta_{\rm ce}$ yields $\Delta_{\rm cs}= 0.0082 \pm 0.0003$~cycle.

\begin{figure}
\centerline{\psfig{figure=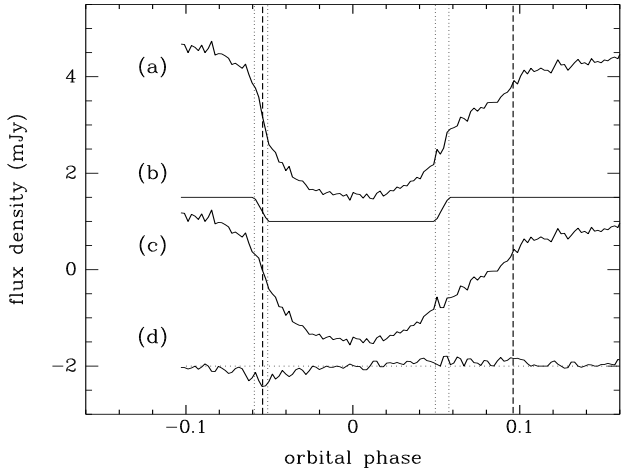,angle=-90,width=9.5cm,rheight=7cm}}
 \caption{ Measuring the eclipse phases of the bright spot.
 (a) Lightcurve 8007. (b) The reconstructed central source lightcurve,
 shifted upwards by 1 mJy. (c) Lightcurve (a) after subtraction of the
 central source component, shifted downwards by 3 mJy. (d) The derivative
 of lightcurve (c), multiplied by a factor 2 and shifted downwards by 2 mJy.
 Dashed lines mark the mid-ingress and mid-egress phases of the bright
 spot and dotted lines mark the four contact phases of the compact
 central source. } 
\label{fig6}
\end{figure}

BS ingress/egress phases ($\phi_{bi},\phi_{be}$) were measured from
the lightcurves in quiescence in which it was possible to simultaneously
identify the eclipse of BS and the egress of CS. We measured the CS
contact and mid-egress phases and used the derived value of $\Delta\phi$
to reconstruct the lightcurve of CS assuming that the flux and duration
of its ingress feature are the same as in egress. Mid-ingress/egress
phases of BS were measured in the lightcurves after removal of the CS
component, which provide and unblended, clean view of the BS ingress
feature (Fig.~\ref{fig6}). 
%
The eclipse parameters measured from the lightcurves in quiescence
are listed in Table~\ref{tab4}. The BS eclipse in lightcurve 7569 starts
earlier and ends later than in the other lightcurves, indicating a
relatively larger disc radius at this epoch (see section~\ref{geom}).
%
\begin{table*}
 \centering
 \begin{minipage}{95mm}
  \caption{Eclipse parameters from lightcurves in quiescence.} \label{tab4}
  \begin{tabular}{@{}lcccccc@{}}
cycle & $\phi_{bi}$ & $\phi_{be}$ & $\phi_{c3}$ & $\phi_{c4}$ &
$\phi_{ce}$ & $\Delta_{ce}$ \\ [1ex]
7569 	& $-0.061$ & +0.099 & +0.0480 & +0.0620 & +0.0550 & 0.014 \\
8007 	& $-0.055$ & +0.096 & +0.0495 & +0.0575 & +0.0540 & 0.008 \\
8008 	& $-0.055$ & +0.096: & +0.0490 & +0.0600 & +0.0550 & 0.011 \\
8012	& $-0.056$ & +0.095 & +0.0490 & +0.0600 & +0.0545 & 0.011 \\
8016+8017 & $-0.054$ & +0.096 & +0.0510 & +0.0580 & +0.0540 & 0.007 \\
\end{tabular}
\end{minipage}
\end{table*}

\subsubsection {Mass ratio, inclination and disc radius} \label{geom}

Making the usual assumption that the secondary star fills its Roche
lobe and given the duration of the eclipse of the central parts of
the disc, $\Delta\phi$, there is a unique relation between the mass
ratio $q= M_2/M_1$ and the binary inclination $i$ (Bailey 1979;
Horne 1985). From Table\,\ref{tab3}, the width of the eclipse in
EX~Dra is $\Delta\phi= 0.1085$. This gives the constraint $q>0.64$,
with $q=0.64$ if $i=90\degr$.

\begin{figure}
\centerline{\psfig{figure=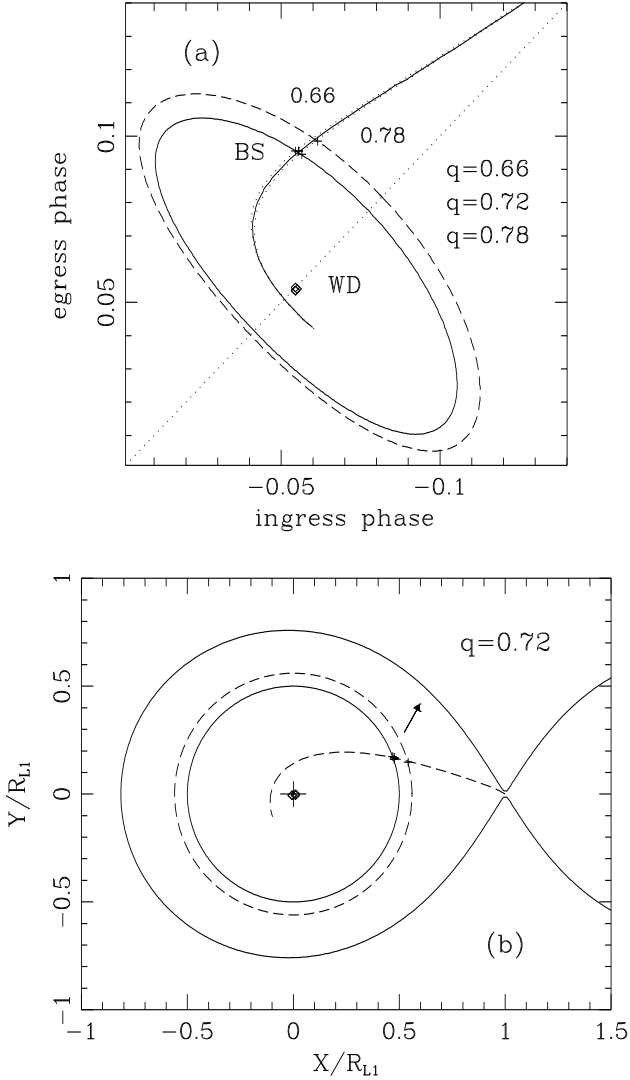,width=11.5cm,rheight=15cm}}
 \caption{ Inferring the binary geometry from the ingress/egress phases
 of CS and BS. (a) Ingress-egress phases diagram. The observed phases
 of mid-ingress/egress of CS are marked with diamonds, those of BS
 with crosses. A diagonal dotted line depicts the line joining the
 component stars. Theoretical gas stream trajectories for three values
 of $q$ are plotted. The stream of matter passes through the position
 of BS for $q=0.72$. The squashed circles represent the accretion discs
 whose edges pass through the two distinct positions of the BS. For
 $q=0.72$, this yields disc radii of $R_{bs}= 0.50$ and $0.56\;
 R_{L1}$. (b) The adopted geometry of the binary for $q=0.72$.
 The observed positions of CS and BS are shown with the theoretical
 gas stream trajectory and discs of radii $R_{bs}= 0.50$ and $0.56\;
 R_{L1}$. The direction at which the bright spot is at maximum is
 indicated by an arrow. } 
\label{fig7}
\end{figure}

When combined with the measured eclipse phases of the CS and BS, this
relation gives a unique solution for $q$, $i$, and $R_{bs}/R_{L1}$,
where $R_{bs}$ is the distance from disc centre to the BS (usually taken
to be the disc radius) and $R_{L1}$ is the distance from disc centre
to the inner lagrangian point L1 (e.g., Smak 1971; Cook \& Warner 1984). 
Fig.\,\ref{fig7}(a) shows a diagram of ingress versus egress phases for
the measurements of the CS and BS in Tables~\ref{tab3} and \ref{tab4}.
Measurements of the CS ingress and egress are shown as the cluster of
small diamonds around phases ($-0.054, +0.054$) in the lower portion of
the diagram. Eclipse phases of BS are indicated by crosses.
%
Theoretical gas stream trajectories corresponding to a set of pairs 
$(i,q)$ are also shown. The trajectories were computed by solving the
equations of motion in a coordinate system synchronously rotating with
the binary, using a 4th order Runge-Kutta algorithm (Press et al. 1986)
and conserving the Jacobi integral constant to one part in $10^{6}$.
The correct mass ratio, and hence inclination, are those for which the calculated stream trajectory passes through the observed position of
the bright spot. 
This yields $q= 0.72 \pm 0.06$ and $i= 85^{+3}_{-2}$ degrees, where the
uncertainties are taken from the standard deviation of the points about
the trajectory of best fit.  Fig.~\ref{fig7}(b) shows the geometry of
the binary system for $q=0.72$. For this mass ratio, the relative size
of the primary Roche lobe is $R_{L1}/a= 0.534\pm 0.009$, where $a$ is
the orbital separation.

BS eclipse phases are clustered at two distinct positions along the
best-fit stream trajectory. The squashed circles in Fig.~\ref{fig7}(a)
represent the accretion discs whose edges pass through these positions
for the adopted mass ratio.  This corresponds to disc radii of 
$R_{bs}/R_{L1}= (0.50\pm 0.01)$ and $(0.56 \pm 0.01)$ with the bright
spot making angles of, respectively, $\alpha_{bs}= 20\degr \pm 1\degr$
and $16\degr \pm 1\degr$ with respect to the line joining both stars.
Circles with these radii are depicted in Fig.~\ref{fig7}(b).
The larger disc radius comes from measurements of BS phases just after
the end of an outburst (lightcurve 7569, end of outburst A) while the
remaining points correspond to lightcurves well into quiescence
(lightcurves 8007, 8008, 8012, 8016+8017). This result suggests that
the accretion disc of EX~Dra shrinks (by at least $\simeq 12$ per cent)
during quiescence -- a similar behaviour to that found in other dwarf
novae (e.g., Smak 1984, 1991; Wood et~al. 1989).
The calculated radii are a factor of 2--3 larger than the radius expected
for zero-viscosity discs, $R_d/R_{L1}\!=\!0.19$ (Flannery 1975), but 
are smaller than the radius expected for pressureless discs,
$R_d/R_{L1}\!=\!0.66$ (Paczy\'{n}ski 1977).

Lightcurve 8074 gives the best phase coverage of the orbital hump in
our dataset. The hump can be well described by a sinusoid of amplitude
0.6 mJy and maximum at orbital phase $-0.17\pm 0.01$ cycle. The 
direction of hump maximum (i.e., maximum visibility of the bright spot)
is indicated in Fig.~\ref{fig7}(b) by an arrow; it is clearly different
from the radial direction of the bright spot. If the hump maximum is
normal to the plane of the shock at the bright spot site then the
shock lies in a direction between the stream trajectory and the edge
of the disc, making an angle of $41\degr \pm 4\degr$ with the latter.

\subsubsection{Masses and radii of the component stars}

An estimate of the binary parameters of EX~Dra may be obtained by
combining the inferred mass ratio $q$ with the empirical main sequence
mass-radius relation of Smith \& Dhillon (1998),
\begin{equation}
R_2/R_\odot = \alpha\;\left( M_2/M_\odot \right)^\beta \;\; ,
\label{zams}
\end{equation}
where $\alpha= 0.91 \pm 0.09$ and $\beta= 0.75\pm 0.04$. The latter 
assumption seems reasonably well justified by the good quality of the
fit to the measured masses of secondary stars in CVs and of field main
sequence stars.

\begin{figure*}
\centerline{\psfig{figure=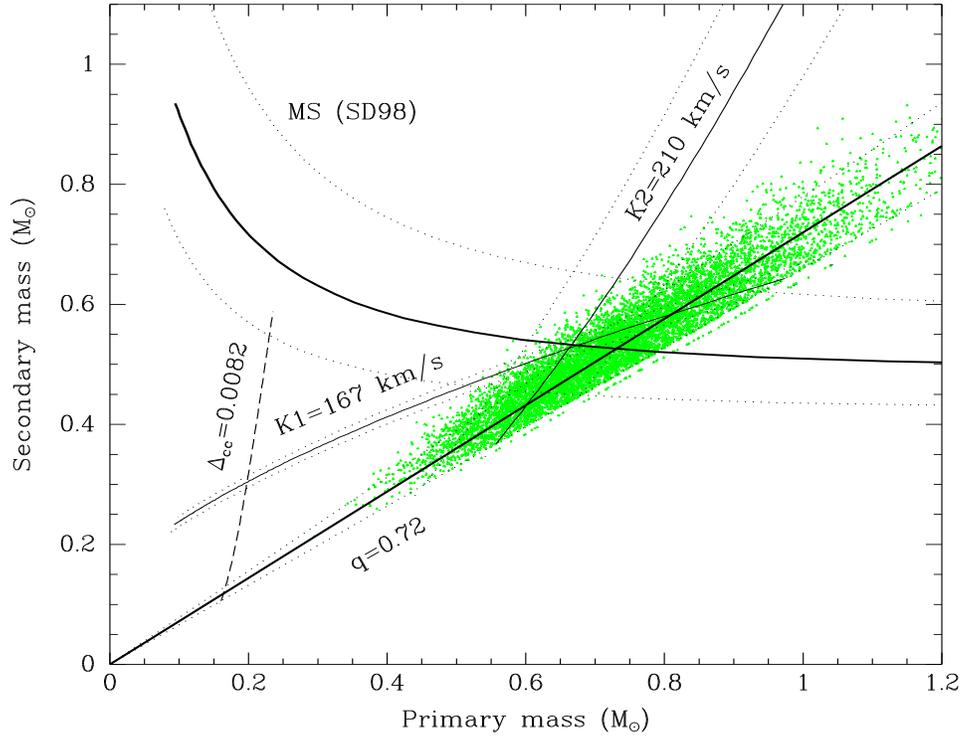,angle=-90,width=15.3cm,rheight=11cm}}
 \caption{ Primary-secondary star mass diagram for EX~Dra. Thick solid
 lines show the constraints obtained from the inferred mass ratio of
 $q=0.72$ and the empirical mass-radius relation of Smith \& Dhillon 
 (1998) [SD98]. Solid lines illustrate the mass functions for a white dwarf
 radial velocity of $K_1= 167\;km\,s^{-1}$ (Fiedler et~al. 1997) and
 a radial velocity of the secondary star of $K_2= 210\;km\,s^{-1}$
 (Billington et~al. 1996). Dotted lines indicate the 1-$\sigma$ limit
 on these relations. The gray cloud of points shows the confidence
 region and is the result of a $10^4$ points Monte Carlo simulation with
 the value of $q$ and the coefficients of the empirical mass-radius
 relation. } 
\label{fig8}
\end{figure*}

The primary-secondary mass diagram for EX~Dra can be seen in 
Fig.~\ref{fig8}. The constraints from the mass ratio and the empirical
mass-radius relation are shown as thick solid lines. We also plotted
lines corresponding to the mass functions for a radial velocity of
the white dwarf of $K_1= 167\; km\,s^{-1}$ (Fiedler et~al. 1997) and
for a radial velocity of the secondary star of $K_2= 210\;km\,s^{-1}$
(Billington et~al. 1996). The four relations are consistent at the
1-$\sigma$ level. 
%
A Monte Carlo propagation code was used to estimate
the errors in the calculated parameters.  The values of the input
parameters $q$ and ($\alpha,\beta)$ are independently varied according
to Gaussian distributions with standard deviation equal to the 
corresponding uncertainties. The results, together with their 1-$\sigma$
errors, are listed in Table~\ref{tab5}.
%
\begin{table}
 \centering
 \begin{minipage}{84mm}
  \caption{Comparison of binary parameters.} \label{tab5}
  \begin{tabular}{@{}lcccc@{}}
parameter & \multicolumn{4}{c}{reference} \\ [-0.5ex]
& this work & (1) & (2) & (3) \\ [1ex]
$q$			& $0.72\pm 0.06$ & 0.80 & 0.75 & 0.73-0.97 \\
$i$			& $85\degr$ ($+3\degr$/$-2\degr$) & 84.1 & 84.2 & 82.1 \\
$M_1/M_\odot$	& $0.75\pm 0.15$ & 0.66 & 0.75 & 0.70 \\
$M_2/M_\odot$	& $0.54\pm 0.10$ & 0.52 & 0.56 & 0.59 \\
$R_1/R_\odot$	& $0.011\pm 0.002$ & 0.011 & 0.013 \\
$R_2/R_\odot$	& $0.57\pm 0.04$ & & 0.57 & 0.59 \\
$a/R_\odot$	& $1.61\pm 0.10$ & & 1.63 & 1.58 \\
$R_d/a$ (quies.) & $0.267\pm 0.004$ & 0.21 & 0.27 \\
$\alpha_{bs}$ (quies.) & $20\degr \pm 1\degr$ \\
$R_{L1}/R_\odot$  & $0.85\pm 0.04$ & & 0.85 & 0.82 \\
$K_1\; (km\;s^{-1})$ & $163\pm 11$ & 167 & 167 & 176 \\
$K_2\; (km\;s^{-1})$ & $224\pm 17$ & 210 & 223 & 210 \\
$v_2\,\sin i \;(km\;s^{-1})$ & $136\pm 9$ & 140 & & 140 \\ [0.5ex]
\multicolumn{5}{l}{(1)= Billington et~al. (1996), (2)= Fiedler et~al. 
(1997),} \\
\multicolumn{5}{l}{(3)= Smith \& Dhillon (1998).} \\
\end{tabular}
\end{minipage}
\end{table}
The quoted $K_1, K_2$ and
$v\sin i$ are the predicted values of, respectively, the radial velocity
of the primary and secondary stars and the secondary star rotational
velocity; $a$ is the binary separation, and the remaining parameters
are self-explanatory. The cloud of points in Fig.~\ref{fig8} was 
obtained from a set of $10^4$ trials using this code. The highest
concentration of points indicates the region of most probable solutions.
Table~\ref{tab5} also lists the estimated parameters of Billington 
et~al. (1996), Fiedler et~al. (1997) and Smith \& Dhillon (1998).
Our model of the EX~Dra binary is in reasonably good agreement with
the models independently derived by those authors from spectroscopic
measurements.

We now turn our attention to the compact central source. An estimate
of its diameter can be obtained from the duration of its ingress/egress
feature in the lightcurve, $\Delta_{cs}$, using the approximate relations
(Ritter 1980),
\begin{equation}
R_{cs}/a= \pi z(q)\; \Delta_{cs} \; \sin\theta \; ,
\;\;\;\;\; \cos\theta= \frac{a}{R_2} \cos i \; ,
\label{larg}
\end{equation}
where $R_{cs}$ and $R_2$ are the radii of the central source and of
the secondary star, respectively, and $z(q)$ is the distance (in units
of $a$) from disc centre to the point tangent to the surface of the
secondary that marks the beginning/end of the eclipse of CS (Baptista 
et~al. 1989).  $z(q)$ is a slow varying function and is usually close
to unity. In our case, for $q=0.72$ we have $z=0.924$.

For the following exercise we adopted the value of $\Delta_{cs}= 0.0082$
cycle inferred from the lightcurves of the low state (section~\ref{phases}).
We note that the width of the CS ingress/egress feature is usually larger
than this in the lightcurves in quiescence (see Table~\ref{tab4});
from the mean $B$ lightcurve in quiescence of Fiedler et~al. (1997) 
(see their fig.\,7) we estimate a width of the egress feature of 0.010
cycle.

Assuming that the compact central source is the white dwarf, the
substitution of Kepler's third law into equation (\ref{larg}) yields a
relation between the mass and the radius of the white dwarf than can be
combined with the Hamada-Salpeter (1961) mass-radius relation (c.f.
Nauenberg 1972) to eliminate $R_{cs}$ and solve for $M_1(q,\Delta_{cs})$
(e.g., Baptista et~al. 1998). This relation (plotted in Fig.~\ref{fig8}
as a dashed line) predicts unreasonably low white dwarf masses of $M_1
\simeq 0.2\; M_\odot$, in clear disagreement with the results in
Table~\ref{tab5}. The discrepancy is not alleviated by the use of a
mass-radius relation for hot white dwarfs (Koester \& Sh\"onberner 1986;
Vennes, Fontaine \& Brassard 1995), the inclusion of possible spherical
distortion effects due to fast rotation of the white dwarf or the 
consideration of strong limb-darkening effects (Wood \& Horne 1990), 
or by adopting the larger value of $\Delta_{cs}$ from the quiescent 
lightcurves. The measured $\Delta_{cs}$ is simply too large for a 
$\simeq 0.7 \;M_\odot$ white dwarf in a binary with the orbital period
of EX~Dra. Together with the variability in flux and duration of the
ingress/egress feature this leads to the conclusion that the observed
compact central source is not a bare white dwarf.

The substitution of the parameters of Table~\ref{tab5} into equation
(\ref{larg}) yields $R_{cs}(\Delta_{cs}= 0.0082)= 0.23\;a = 0.037\; 
R_\odot = 3.36\; R_1$. Therefore, the white dwarf in EX~Dra seems
surrounded by an extended, variable atmosphere or boundary layer of at
least 3 times its radius. In this regard EX~Dra is similar to the long
period, eclipsing dwarf nova IP~Peg -- where the white dwarf seems to
be wrapped in a thick boundary layer more than twice its radius (Wood
\& Crawford 1986) -- and is clearly different from the short-period
dwarf novae OY~Car, Z~Cha and HT~Cas, where the central source seems
to be a bare white dwarf (Wood \& Horne 1990). This result suggests
that different physical conditions may exist in the inner disc regions
of the short period and of the long period dwarf novae, possibly related
to the distinct mass accretion rates of these groups, although the
statistics of eclipsing dwarf novae is still very low on both sides of
the CV period gap. From the derived parameters, we predict a duration
of the ingress/egress of the white dwarf of $\Delta_{wd}= 0.0024 \pm
0.0006$ cycle.

\subsection {Distance estimates}

The flux densities at mid-eclipse of the flat-bottomed lightcurves of
the low state (Fig.~\ref{fig3}) yield an upper limit to the contribution
of the secondary star that can be used to set a lower limit on the
distance to EX~Dra (e.g., Baptista, Steiner \& Cieslinski 1994).

We find $F_{\rm mid}(V)= 0.79\pm 0.05$ mJy and $F_{\rm mid}(R)= 1.56 \pm
0.03$ mJy, where the quoted values are the median of the fluxes in the
phase range ($-0.02,+0.02$) cycle and the uncertainties were derived from
the median of the absolute deviations with respect to the median. This
corresponds to an apparent magnitude of $V_{\rm mid}= 16.66\pm 0.07$ mag
and a color index of $(V-R)= +0.92\pm 0.07$ mag. We estimated a reddening
of $E(B-V)= 0.15\; {\rm mag}\;kpc^{-1}\; (A_v= 4.8\times 10^{-4}\; {\rm
mag}\;pc^{-1})$ for EX~Dra from the galactic interstellar extinction contour
maps of Lucke (1978), which gives a color excess of $E(V-R)= 0.02$ mag 
for a distance of $D=290\;pc$ (see below). This leads to a corrected,
intrinsic color index of $(V-R)_0= +0.90\pm 0.07$ mag.

\begin{figure}
\centerline{\psfig{figure=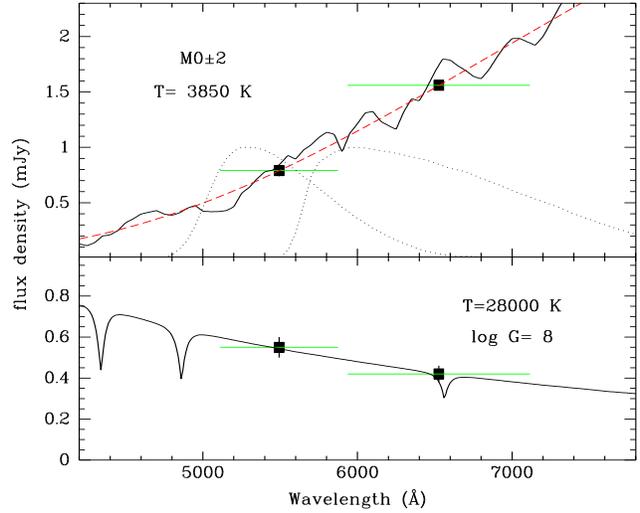,angle=-90,width=10cm,rheight=7cm}}
 \caption{ Top: $V$ and $R$ mid-eclipse fluxes (open squares) and best-fit
 main sequence (solid) and blackbody (dashed) spectra. Dotted curves
 show normalized response functions of the $V$ and $R$ passbands. 
 Horizontal bars show the calculated flux of each model in the 
 corresponding passband and marks the full-width half-maximum of each
 passband. Bottom: the measured fluxes of the central source and the 
 best-fit white dwarf atmosphere model. The notation is the same as above. } 
\label{fig9}
\end{figure}

A main sequence star with this color index has a spectral type 
M$0\pm 2$, an absolute magnitude of $M_v= 9.2\pm 0.6$ mag and an 
effective surface temperature of $T_{\rm eff}= 3850\pm 200\;K$ 
(Schmidt-Kaler 1982; Pickles 1985). This spectral type is in good
agreement with that inferred from the spectroscopy by Billington et~al.
(1996) and Fiedler et~al. (1997). With the assumption that the observed
properties of the secondary star in EX~Dra are similar to those of a
normal main sequence star of same mass, we replace these values into
the equation,
\begin{equation}
5\,\log D(pc)= V_{\rm mid} - M_v + 5 - A_v\, D(pc) \;\;\; ,
\end{equation}
to find a photometric parallax distance of $D_{MS}= (290\pm 80)\;pc$, where
the quoted uncertainty is obtained from the propagation of errors in the 
input parameters and do not account for possible systematic errors.
If we neglect the interstellar extinction, the inferred distance is
reduced to $D=280\pm 80\;pc$. An alternative blackbody fit to the
mid-eclipse fluxes including the interstellar extinction yields a
solid angle of $\theta^2_{BB}= [(R_2/R_\odot)/(D/kpc)]^2= 3.8\pm 0.8$,
an $T_{\rm eff}= 3550\pm 250\;K$, and a distance of $D_{BB}= (290\pm
40)\;pc$. The measured mid-eclipse fluxes and the best-fit main sequence
and blackbody spectra are shown in the upper panel of Fig.~\ref{fig9}.
%

Another distance estimate can be obtained from the $V$ and $R$ flux
densities of the compact central source in the lightcurves of the low
state, since we also have an estimate of the physical dimension of this
source. The fluxes at egress, which are free from contamination of light
from the bright spot, were obtained by integrating the derivative
between the third and fourth contact phases (see section~\ref{phases}).
We find $F_{cs}(V)= 0.55\pm 0.05$ mJy and $F_{cs}(R)= 0.42\pm 0.04$ mJy.
We fitted the observed flux densities from synthetic photometry with
white dwarf atmosphere models (G\"ansicke, Beuermann \& de Martino 1995)
allowing for the effect of interstellar extinction as estimated above. 
The best-fit model has $T_{cs}= (28\pm 3)\times 10^3\;K,\:\log g= 8$ 
and $ \theta^2_{cs}= (3.3\pm 0.5) \times 10^{-3}$. The measured central
source fluxes and best-fit white dwarf atmosphere model are shown in
the lower panel of Fig.~\ref{fig9}.

The corresponding distance depends on the geometry and effective emitting
area of the central source as seen by an observer on Earth. A spherical
central source has an effective area of $A_{sp}= \pi R^2_{cs}$
as projected onto the plane of the sky. In this case, if the inner disc
is optically thin, both hemispheres of the central source are seen 
and a distance of $D= 640\pm 50\;pc$ is obtained. If the inner disc is
opaque the lower hemisphere of the central source is occulted and the
distance is reduced to $D= 450\pm 40\;pc$. Both values are in agreement
with the lower limit derived from the contribution of the secondary star.

Alternatively, if we assume that the distance is $D=290\;pc$, the
inferred $\theta^2_{cs}$ allow us to constrain the geometry of the central
source. At this distance the effective area of the central source is
reduced to 21 per cent of $A_{sp}$. Since the equatorial diameter of
the central source is set by the width of its ingress/egress feature
to be $2 R_{cs}$, this implies that the polar diameter is significantly
smaller than $R_{cs}$. Hence, the distance estimate from the flux of
the secondary star and that from the flux of the central source can be
reconciled if the central source has a toroidal shape with an equatorial
diameter of $2\times 0.037\;R_\odot$ and a vertical thickness of $0.012\;
R_\odot$ (if the inner disc is optically thin) or $0.024\;R_\odot$ (if
the inner disc is opaque).

\section{Summary} \label{final}

The results of the analysis of $V$ and $R$ high speed photometry of 
EX~Dra in quiescence and through outburst can be summarized as follows:

\begin{enumerate}

\item During the period of the observations EX~Dra showed outbursts
with typical amplitudes of $\simeq 2.0$ mag, duration of $\simeq 10$
days, and average time between outbursts of $20\pm 3$ days. The observed
amplitudes are larger than those found by Billington et~al. (1996).

\item The lightcurves during outburst were grouped by outburst phase.
The analysis of these lightcurves indicates that the outbursts do not
start in the outer disc regions and, therefore, favours the disc 
instability model. The disc expands during the rise to maximum (as
indicated by the increasing width of the eclipse) and shrinks during
decline. The decrease in brightness at the later stages of the outburst
is due to the fading of the light from the inner disc regions.

\item At the end of two outbursts the system was seen to go through a
phase of lower brightness (named the low state), characterized by the
fairly symmetric eclipse of a compact source at disc centre with little
evidence of a bright spot at disc rim, and by an out-of-eclipse level
$\simeq 15$ per cent lower than the typical quiescent level.

\item New eclipse timings were measured from the lightcurves in
quiescence and a revised ephemeris was derived. The residuals with
respect to the linear ephemeris show a clear cyclical behaviour and
can be well described by a sinusoid of amplitude 1.2 minutes and period
$\simeq 4$ years. This period variation is possibly related to a
solar-like magnetic activity cycle in the secondary star.

\item Eclipse phases of the compact central source and of the bright
spot were used to derive the geometry of the binary. By constraining
the gas stream trajectory to pass through the observed position of the 
bright spot we find $q=0.72\pm 0.06$ and $i= 85^{+3}_{-2}$ degrees.

\item The binary parameters were estimated by combining the measured
mass ratio with the assumption that the secondary star in EX~Dra
obeys the empirical main sequence mass-radius relation of Smith \&
Dhillon (1998). The set of derived parameters in listed in 
Table~\ref{tab5}.

\item The observed changes in the position of the bright spot with
time suggest that the accretion disc shrinks during quiescence by at
least $\simeq 12$ per cent.

\item The phase of hump maximum is distinct from the radial direction
of the bright spot. If the hump maximum is normal to the plane of the
shock at the bright spot site then the shock lies in a direction between
the stream trajectory and the edge of the disc, making an angle of
$41\degr\pm 4\degr$ with the latter.

\item The white dwarf seems surrounded by an extended, variable atmosphere
or boundary layer of at least 3 times its radius. From the derived
parameters, a duration of the ingress/egress of the white dwarf of
$\Delta_{wd}= 0.0024\pm 0.0006$ cycle is predicted.

\item The fluxes at mid-eclipse of the lightcurves of the low state
yield an upper limit to the contribution of the secondary star and lead
to a lower limit photometric parallax distance of $D_{MS}= 290\pm 80\;pc$.

\item The fluxes of the central source are well fitted by a white
dwarf atmosphere model with $T_{cs}= (28\pm 3)\times 10^3\;K ,\: \log g
= 8$ and solid angle $\theta^2_{cs}= [(R_{cs}/R_\odot)/(D/kpc)]^2=
(3.3\pm 0.5) \times 10^{-3}$. For a spherical central source, this leads
to a distance of $D= 640\pm 50\;pc$ if the inner disc is optically thin.
The distance estimates from the mid-eclipse fluxes and from the fluxes
of the central source can be reconciled if the central source has a
toroidal shape with an equatorial diameter of $2\times 0.037\;R_\odot$ 
and a vertical thickness of $0.012\; R_\odot$ (if the inner disc is
optically thin) or $0.024\;R_\odot$ (if the inner disc is opaque).

\end{enumerate}

The analysis of the set of lightcurves through outburst with eclipse
mapping techniques yields an uneven opportunity to investigate the
changes in the structure of an outbursting accretion disc and is the
subject of another paper (Baptista \& Catal\'an 1999, 2000).

\section*{Acknowledgments}

We are grateful to Yvonne Unruh for valuable help in obtaining the data
at JGT and to Boris G\"ansicke for kindly providing the white dwarf 
atmosphere models.
In this research we have used, and acknowledge with thanks, data from
the AAVSO International Database and the VSNET that are based on
observations collected by variable star observers worldwide. 
RB acknowledges financial support from CNPq/Brazil through grant no.
300\,354/96-7. MSC acknowledges financial support from a PPARC
post-doctoral grant during part of this work. This research was
partially supported by PRONEX grant FAURGS/FINEP 7697.1003.00.

\bsp

\end{document}